\documentclass[11pt]{article}

\usepackage[pdftex]{graphicx}
\usepackage{url}
\usepackage{epic}
\usepackage{eepic}
\usepackage{fancybox}
\usepackage{here}
\usepackage{soul}

\begin{document}

\title{A Balance for Fairness: Fair Distribution Utilising Physics in Games of Characteristic Function Form}

\author{
Song-Ju Kim${}^{\dag}$${}^{\ddag}$${}^{\ast}$, Taiki Takahashi${}^{\S}$, and Kazuo Sano${}^{\dag}$${}^{\P}$\\
\\

${}^{\dag}$ SOBIN Institute, Kawanishi, Japan\\
https://sobin.org\\
${}^{\ddag}$ Graduate School of Media and Governance, Keio University, Fujisawa, Japan\\
${}^{\S}$ Department of Behavioral Science, Research and Education Center for Brain Sciences,\\
Center for Experimental Research in Social Sciences, Hokkaido University, Sapporo, Japan\\
${}^{\P}$ Department of Economics, Fukui Prefectural University, Fukui, Japan\\
${}^{\ast}$Email: kim@sobin.org
}

\maketitle

\abstract
{\bf 
In chaotic modern society, there is an increasing demand for the realization of true 'fairness'.
In Greek mythology, Themis, the 'goddess of justice', has a sword in her right hand to protect  society from vices, and a 'balance of judgment' in her left hand that measures good and evil.
In this study, we propose a fair distribution method 'utilising physics' for the profit in games of characteristic function form. 
Specifically, we show that the linear programming problem for calculating 'nucleolus' can be efficiently solved by considering it as a physical system in which gravity works.
In addition to being able to significantly reduce computational complexity thereby, we believe that this system could have flexibility necessary to respond to real-time changes in the parameter.
}
\endabstract


\newpage

\section{Introduction}

The Buddha taught that the 'goodness' that embodies 'righteousness' stands in contrast to duties and to personal ties. 
In other words, the definition of 'good' is what will be in the interest of oneself, in the interest of others, and in the interest of those to be born in the future.
Humans feel happy when they are needed within a community and when they play a role that benefits others and the whole community. 
This core aspect of human nature is often forgotten in actual social activities, seen as a 'beautiful thing' at a distance from the practical. 

In modern society driven by neoliberalism, only the more primitive desires of man are judged essential.
Because neoliberalism assumes that individual negligence are the source of poverty, disparity and poverty are the incentives for hard work rather than goodness or spiritual insight~\cite{wakati}. 
We have come to believe that market-based competition brings efficiency and an optimization of society. 
As a result of this, not only the destruction of the natural environment, but also the destruction of the human environment has been promoted as a means of achieving growth. 
The result has been a loss of trust and bonds between humans due to excessive competition. 

Today, humanity is facing a serious crisis not only because of the COVID-19, but also because of the lack of trust between people. In order to defeat infectious diseases, people need to trust scientists, citizens need to trust public institutions, and countries need to trust each other. 
New viruses can occur everywhere, and if they do occur, the important question will be whether the government can immediately implement a city blockade. Unfortunately, we must include the economic loss of the blockade as a variable. In such a case, the timing of the blockade may be delayed without international assistance~\cite{harari}.

Nevertheless, it has been proven in socialist countries that easy solutions to poverty, such as equal distribution, do weaken incentives for hard work. Also, the fact that acting alone and selfishly within an altruistic community can produce great benefits for the individual (in the short term) seems to be an obstacle to formulating a powerful critique of neoliberal ideology. 
The question that remains is whether it not possible to build a system that mobilizes digital and new analog technologies so that people can do 'good' with confidence?

The problem of 'uneven distribution of wealth'~\cite{pikety}, has been accelerated in the midst of the COVID-19 crisis~\cite{pastreich,pastreich2}. 
The 'universal basic income', which includes wealth redistribution, solving social security problems, and combating poverty, is being actively discussed and experimented with. This year Spain became the first country in the world to implement a 'basic income' for the needy. 
In this chaotic 'new turn of capitalism', there is an urgent need to establish a firm and blameless 'fairness' in order to avoid inequities and social justice for everyone. 
However, there are different kinds of difficulties for the realization as follows.

\begin{enumerate}
\item
Fairness requires proper judgment and the processing of various factors impacting the current situation.
The calculation often cannot keep up with the changing dynamic system. 
\item
The elements critical to determining fairness that depend on human interpretation are difficult to formalize.
\item
As Arrow's impossibility theorem~\cite{arrow,kime} shows, we do not even know whether there is a 'fair solution' or not.
\end{enumerate}

In this study, we propose a method to realize a 'fair distribution of profit' in a physical sense within the restricted domain of games. 
Previously, Kim proposed a flexible analog computation scheme that overcomes the computational difficulties of digital computing by taking advantage of natural (physical) properties, such as conservation laws, continuity, and fluctuations~\cite{NI,bombe}. 
In particular, he extracted an efficient decision-making scheme (reinforcement learning) called the 'tug-of-war principle'~\cite{tow} and created 'self-decision-making physical systems' by applying it to quantum dots~\cite{qdot}, single photons~\cite{single}, atomic switches~\cite{atom}, semiconductor lasers~\cite{laser}, ionic devices~\cite{ion}, resistive memories~\cite{rand}, and IoTs~\cite{IoT}. 

\begin{figure}[H]
 \begin{center} 
  \includegraphics[width=50mm]{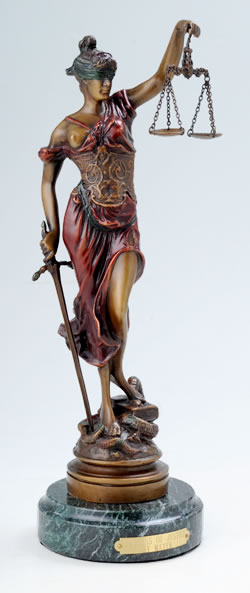}
  \caption{Goddes Themis. This picture is borrowed from the following website.\\ http://www.catalog-shopping.co.jp}
  \label{fig:1}
 \end{center}
\end{figure}

The physics-based computations are useful in the calculation of 'fairness.' 
Themis, the Greek goddess of justice, installed in courthouses and law offices, is considered the 'guardian of justice and order' (see Figure \ref{fig:1}). 
She wears a blindfold as a sign of her objectivity and selflessness, holding a sword in her right hand to protect society from vice and a 'balance of justice' in her left hand to measure right and wrong. 
In other words, she leaves the impartial judgment to nature (physical phenomena) rather than perceptions and attachments. 
Although that balance is symbol, the 'use of nature (physics)' is thought to have the effect of eliminating artificiality and gaining consensus (concessions) from the people concerning issues. 

\begin{figure}[H]
 \begin{center}
\includegraphics[width=100mm]{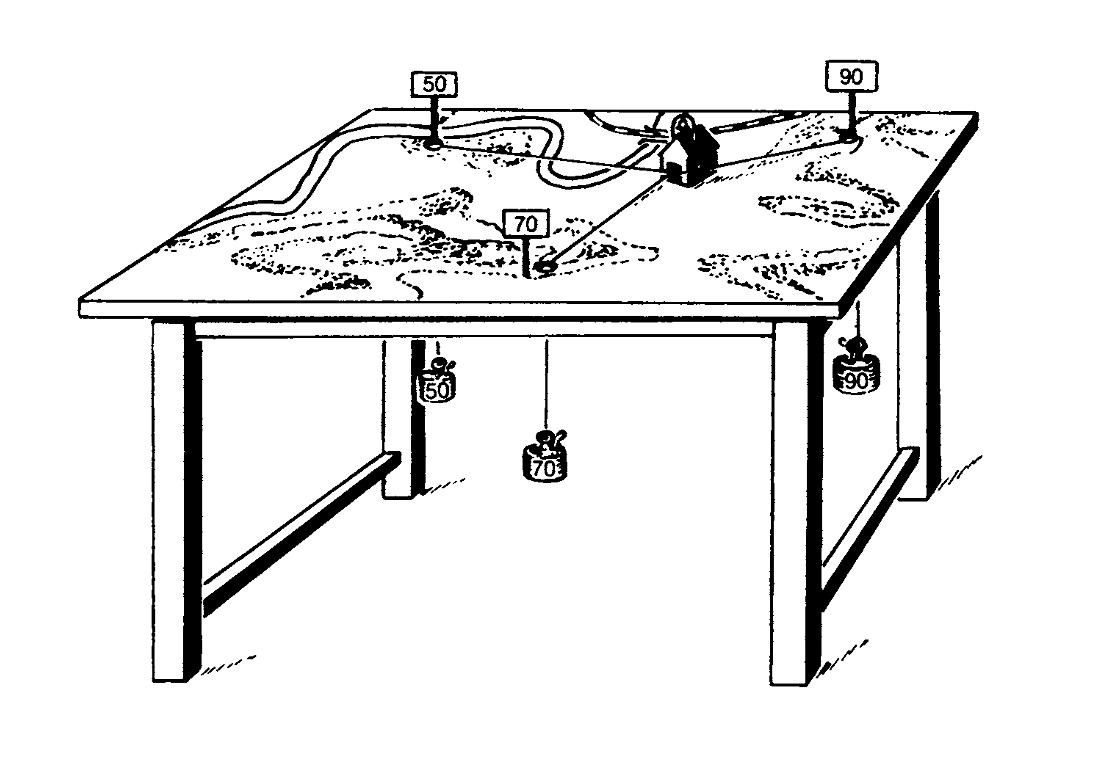}
\caption{An example for physically solving computational problems. This figure is borrowed from the following book.
H. Steinhaus: Mathematical Snapshots, Oxford Univ. Press (1950)} 
\label{fig:2}
\end{center}
\end{figure}

The direct use of physics in computation is not a particularly new concept~\cite{stein}. 
Rather, before the advent of the digital computer, everything was analog~\cite{kikai}. 
For example, suppose that Figure \ref{fig:2} has three villages with 50, 70, and 90 elementary school students living in each village. 
How can we minimize the total distance travelled for all students' to school when we build an elementary school?
As shown in the figure, the center of gravity is the answer (as to where the elementary school should be built) made by hanging weights determined by the number of students in each village. 
If we consider only the possible locations, the calculation would be intractable, however, by 'designing' the problem in this way, we can leave the calculation (or even part of it) to physics.

\section{Games in Characteristic function Forms}

A cooperative game wherein the players can cooperate with each other can be an effective means of determining how the profit obtained through cooperation can be fairly distributed~\cite{tokusei}.
Cooperative games have been applied to market analysis, voting analysis, and to cost sharing analysis in economics, as well as to market design, such as auction theory and matching theory, the theme of the 2020 Nobel Prize in Economics~\cite{tokusei2}.

Let us define $N = \{ 1, 2, \cdots, n \}$ as the set of $n$ players. 
We call the non-empty subset of $N$ 'coalitions'. 
If $R$ is a set of real numbers, we call the real value function $v : 2^N \rightarrow R$  on $2^N$ (the set of the subset of $N$) 'a characteristic function'. 
For each coalition $S$ ( $\subseteq N$ ), $v(S)$ represents the profit that can be gained by the members of the coalition $S$ by cooperating. 
More concretely, it is the difference between the cooperative case and the non-cooperative case that is critical.
 We call $(N, v)$, which is the set of players $N$ and a characteristic function $v$, a 'game of characteristic function form'.
In the following subsections, we introduce two examples of the 'taxi problem' and the 'bankruptcy problem'.

\subsection{Taxi problem}

Consider a situation in which three persons (1, 2, 3) finish eating at a restaurant and go to home by taxi. 
If they return home alone, the charges are person 1: 20\$, person 2: 21\$, and person 3: 25\$, respectively. 
In addition, between each of the two houses, the following taxi fees are charged: 10\$ between 1 and 2, 14\$ between 1 and 3, and 20\$ between 2 and 3. 
In this case, the shortest route is (2 - 1 - 3), which costs 45\$. 

In this example, the characteristic functions are the followings. 
\begin{eqnarray}
v(1, 2, 3) & = & (20 + 21 + 25) - 45 = 21,\\
v(1, 2)  & = & 41 - 30 = 11,\\
v(1, 3) & = & 45 - 34 = 11,\\
v(2, 3) & = & 46 - 41 = 5,\\
v(1)  & = & v(2) = v(3) = 0. 
\label{eq:1}
\end{eqnarray}
where a characteristic function $v(x, y)$ is the profit if $x$ and $y$ are cooperated, that is, the difference between the total amount of fares for going home alone and the fare for going home in a taxi by the shortest route. 

The question is, 'How should the profit of 21\$ be divided 'fairly' among the three? 
In general, this problem can be solved using the concept of 'nucleolus ~\cite{jin}' as follows. 
First, in the games of characteristic function form $(N, v)$, we define the 'complaint' $C(S, x)$ of the coalition $S$ with respect to the allocation $x$ as follows. 
\begin{equation}
C(S, x) = v(S) - \sum_{i \in S} x_i 
\end{equation}
That is, the profit of the coalition $S$' minus the 'total allocation (distribution) of those who participated in the coalition $S$'. 
The 'nucleolus' is the allocation that minimize the largest complaint and is generally known to be uniquely determined (for simplicity, we avoid an exact mathematical definition of 'nucleolus' here). 

In the case of the taxi problem, the complaint against each coalition is as follows 
\begin{eqnarray}
C(\{1,2\},x) & = & v(1,2) - x_1 - x_2  = 11 - x_1 - x_2, \\
C(\{1,3\},x) & = & v(1,3) - x_1 - x_3 = 11 - x_1 - x_3, \\
C(\{2,3\},x) & = & v(2,3) - x_2 - x_3 = 5 - x_2 - x_3,\\
C(\{1\},x) & = & v(1) - x_1 = -x_1, \label{c1}\\
C(\{2\},x) & = & v(2) - x_2 = -x_2 , \label{c2}\\
C(\{3\},x) & = & v(3) - x_3 = -x_3 .\label{c3}
\end{eqnarray}
Using the total rationality $x_1 + x_2 + x_3$$ = $$v(1,2,3)$$ = $$21$, we rewrite the complaints of the two-person coalitions.
\begin{eqnarray}
C(\{1,2\},x) & = & x_3 - 10,\label{c4}\\
C(\{1,3\},x) & = & x_2 - 10,\label{c5}\\
C(\{2,3\},x) & = & x_1 - 16.\label{c6}
\end{eqnarray}

Since the nucleolus is the allocation that minimizes the maximum constraint, we suppose above six complaints eqs. (\ref{c1}, \ref{c2}, \ref{c3}, \ref{c4}, \ref{c5}, and \ref{c6},) are less than or equal to a certain value $T$, then we minimize $T$.
In other words, consider the following linear programming problem.

$Minimize(T):$ 
\begin{eqnarray}
-T \leq & x_1 & \leq T+ 16,
  \label{eq:z1}\\
-T \leq & x_2 & \leq T + 10,
  \label{eq:z2}\\
-T \leq & x_3 & \leq T + 10,
  \label{eq:z3}\\
  x_1 + x_2 + & x_3 & =  21,
  \label{eq:z}\\
  & x_1 & \ge 0, \\ 
  & x_2 & \ge 0, \\
  & x_3 & \ge 0.
\end{eqnarray}

In order for there to be $x_1, x_2, x_3$ that satisfies the constraints, the following formulas must hold from the eqs. (\ref{eq:z1}, \ref{eq:z2}, \ref{eq:z2}, and \ref{eq:z3}) and the individual rationality ($x_i$$\ge$$v(i)$). 
\begin{eqnarray}
-T  \leq T + 16 &  \Longleftrightarrow  & T \ge -8, \\
-T   \leq T + 10  &  \Longleftrightarrow & T \ge -5.
\end{eqnarray}
Above expressions can be transformed to the followings from eqs. (\ref{eq:z1}, \ref{eq:z2}, and \ref{eq:z3}) and the total rationality eq. (\ref{eq:z}).
\begin{eqnarray}
-3T & \leq & (x_1 + x_2 + x_3) = 21  \leq 3T + 36  \\
& \Longleftrightarrow & T \ge -7   \hspace{2mm}   and \hspace{2mm}  T \ge -5
\end{eqnarray}

The minimum value of $T$ that satisfies all these conditions is $-5$, which, when substituted, yields the following constraint condition, 
\begin{eqnarray}
  -5 \leq &x_1& \leq 11,\\  
  -5 \leq &x_2& \leq 5, \\
  -5 \leq &x_3& \leq 5, \\
  x_1 + x_2 + & x_3 & = 21
\end{eqnarray}

$T = -5$ can be realized only with $x_1 = 11$, $x_2 = 5$, and $x_3 = 5$.
That is, 'nucleolus' in the taxi problem is $(11, 5, 5) $ and 
According to the nucleolus's criterion, person1 would receive 11\$, and 2 and 3 would both receive the allocation of 5\$.
In general, in games of characteristic function form with $n$ players, we must solve a linear programming problem to minimize the maximum complaint and if there is more than one allocation that achieves the smallest maximum complaint, the second largest complaint in the allocation must be minimized.
To find 'nucleolus', it is necessary to solve the linear programming problems at most $n - 1$ times.
The computational complexity of the linear programming is commonly known as at most polynomial time~\cite{tokusei}.

\subsection{Bankruptcy problem (estate distribution problem)}

In what can be considered a game of characteristic function form, there exists a famous problem known since long ago. 
If a person goes bankrupt leaving a debt of $D_1, D_2, \cdots, D_n$ to each of the $n$ creditors $1, 2, \cdots, n$, 
How should the bankrupt's entire estate ($M$) be 'fairly' distributed to each creditor?
At first glance, a proportional division according to the size of the debt seems to be 'fair'. 
However, if we consider a case in which the value of the entire estate is smaller than the amount of any debt we encounter a situation in which, because all creditors are similarly unable to fully recover their share, it is better to split the debt in equal parts. 

In fact, this sort of bankruptcy problem is described in the Babylonian Talmud, a Jewish scripture more than 2,400 years ago (Table \ref{table:1}). 
In Table \ref{table:1}, E represents the total estate and D represents the creditors and the amount of their debts. 
If the total estate is $300$, the division is proportional, but if it is $100$, it is an equal division. 
Even though for $200$, the criteria are not as clear at first glance, there exists a clear criterion of 'nucleolus' which was first formulated in 1969. 
It is very surprising that the Sages of the Talmud more than 2400 years ago had clear ideas and criteria for the concepts of 'justice' and 'fairness' in this specific financial case. 
We had to wait until 1985 for a satisfactory explanation of this criterion of bankruptcy problem~\cite{aumann}. 
That was accomplished by Aumann and Maschler's discussion in relation to game theory~\cite{hasan}.

Aumann and Maschler went on to read from the following description of other passages in the Talmud Mishnah (Baba Metzia 2a) 

\ \\
{\bf
Two hold a garment; one claims it all, the other claims half. 
Two hold a garment; one claims it all, the other claims half.
Then the one is awarded 3/4, the other 1/4.
}
\ \\

This statement says that if there are those who demand the whole thing and those who demand half of it, it should be divided into $3/4$ and $1/4$ respectively. 
This is neither a proportional nor an equal division, but a 'residual equal division'. 
For some reason the latter allows $1/2$ to the former, so the remaining $1/2$ is to be divided equally.  

Consider the case of two creditors. 
The amount that creditor 1 is willing to admit to creditor 2 is the greater of $(M - D_1)$ and $0$, so $(M - D_1)_+$, defined as $x_+ = max(x, 0)$. 
Similarly, the amount that creditor 2 is willing to allow to 1 is $(M - D_2)_+$. 
The shares of creditors 1 and 2 $X_1$ and $X_2$ are as follows 
\begin{eqnarray}
X_1 & = & (M - D_2)_+ + \frac{\{M - (M - D_1)_+ - (M - D_2)_+\}}{2} ,\\
X_2 & = & (M - D_1)_+ + \frac{\{M - (M - D_1)_+ - (M - D_2)_+\}}{2} .
\label{eq:2}
\end{eqnarray}
They called the Contested Garment principle the distribution principle in a bankruptcy case with one bankrupt and two creditors. 
In the Talmudic bankruptcy problem (three creditors) the CG principle holds for any two persons and is shown to be 'nucleolus'~\cite{aumann}.

\section{Physical Method for Fair Distribution}

\begin{figure}[H]
 \begin{center}
\includegraphics[width=100mm]{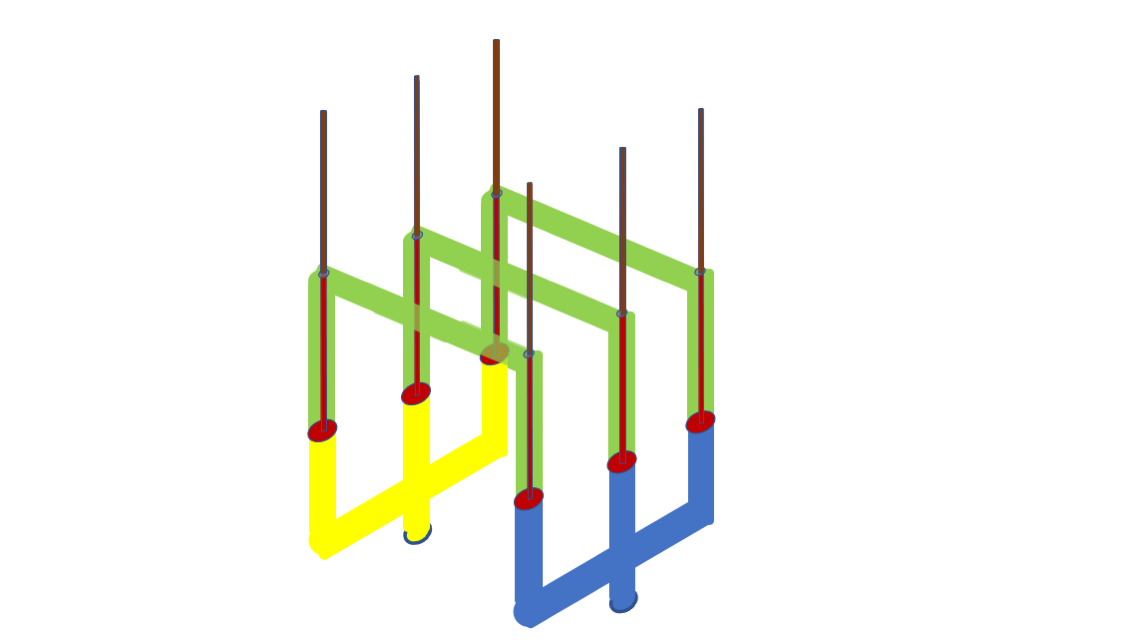}
\caption{An implementation for interlocked adjuster moving. Each cylinder filled with green liquid is connecting blue and yellow cylinders. } 
\label{fig:5}
\end{center}
\end{figure}

In this section, we mainly describe how to solve a game of characteristic function form with 'gravity'. 
First, we have two cylinders A (blue) and B (yellow) filled with a liquid (incompressibility) such as Figure 4. 
The adjusters of A and B are assumed to be interlocked: if one moves, the other moves the same amount in the opposite direction (see Figure  \ref{fig:5} ). 
By moving the adjusters in sequence according to the instructions as in Figure 4, each liquid level represents the 'amount of complaint' of the corresponding coalition. 
The liquid levels $E_1$, $E_2$ and $E_3$ of the blue liquid correspond to the definitions of complaints $C(\{2,3\},x)$, $C(\{1,3\},x)$ and $C(\{1,2\},x)$ of the previous section, respectively.
Finally, when the two cylinders are set up vertically at the same time to exert gravity, we find that gravity sets the liquid level in motion and that the final equilibrium liquid level will represent the 'fair distribution'. 

The procedure for the solution with physics is as follows 
\begin{enumerate}
\item
The valve is open so as to allow for the adjustment of the liquid level by moving the adjuster. 
For a cylinder of blue liquid A, divide equally the profit $v(1,2,3)$ that would be obtained if everyone had cooperated. 
$X_i$ $=$ $v(1,2,3) / 3$.
\item
For each $i$, add the profit of two-person coalition other than $i$ ($ + v(j, k)$), and then 
subtract the profit that would be gained if everyone had cooperated ($ - v(1,2,3)$). 
\item
For each $i$, we record the current liquid level and then set it to the origin of the upcoming gravitational change in the liquid level $y_i$.
That is, each distribution will be $X_i = y_i + v(1,2,3) / 3$, satisfying the requirements of the conservation law $y_1 + y_2 + y_3 = 0$.
The current liquid level $E_i$ expresses the complaint of the coalition $(j, k)$.
That is, $E_i$ $\equiv$ $C(\{j, k\},x)$ $=$ $v(j, k) - (X_j + X_k)$$ = $$X_i -  v(1,2,3) + v(j, k)$
\item
Then, for each $i$ of yellow liquid cylinder B, 
add the profit generated by the coalition $i$ $( + v(i))$, and then subtract  $ - v(1,2,3) / 3$. 
The current liquid level $C_i$ will expresses the complaint of the coalition $i$.
That is, $C_i \equiv v(i) - X_i$.
\item
All six complaints will thereby be expressed and the preparation will be complete. 
At this point, the valves are closed and the adjusters of cylinders A and B are interlocked. 
That is, an increase in cylinder A's $E_i$ results in a corresponding decrease in the corresponding liquid level $C_i$ of the cylinder B.
\item
Stand the two cylinder systems vertically at the same time so that gravity works (see Figure \ref{fig:5} ).
\item
Wait until the equilibrium state is attained and then record the final liquid level change $y_i$. 
\item
Note that the liquid level change satisfies the following equation:
\begin{equation}
y_i  \ge v(i) - v(1,2,3) / 3.
\end{equation}
\item
The distribution is determined by the following equation: 
\begin{equation}
X_i = y_i + v(1,2,3) / 3.
\end{equation}
\end{enumerate}

\section{Physical Solutions on Fair Distribution Problem}

The physical solutions of the taxi problem and the bankruptcy problem (three ways) are shown in Figure \ref{fig:4}. 
We can confirm that we are able to find solutions to all problems. 
There are various possibilities depending on whether the 'interlocking adjusters' in the figure are realized with other liquids or implemented mechanically, 
In any case, at each level of equilibrium, the tensile force and the pressure inside the liquid (the sum of the statistical and gravitational pressures according to Pascal's principle) will come to a balance. 
It is obvious that the 'equal liquid level' of every cylinder will be realized in the equilibrium. 
However, the system cannot reach the ideal equilibrium state due to the constraints of the adjusters in this system.
The highest liquid level (maximum complaint) will gradually decrease, heading towards the equal liquid level.
However, at the same time, other liquid levels will increase due to interlocking, with the result that the highest liquid level will eventually stop when it cannot drop any further.

\section{Summary and Discussion}

In this study, we have shown that the fair distribution of profit in games of characteristic function form can be solved utilising physics. 
More specifically, the linear programming problem used to compute 'nucleolus' can be regarded as a physical system that responds to gravity, and the solution can be found efficiently. 
However, it is only shown to be effective for the solving of relatively simple problems at present.
It is not yet known whether this method for solving by making use of physics can also be effective for more complex problems. 
Identification of the necessary conditions (physical parameters, constraints, etc.) for this physical solution to work is also future works. 

In order for this approach to become a candidate, a viable method, for overcoming the computational difficulties in digital computers, it is necessary to at least address the general $N$-person problems, and how the solution should be presented. 
Nevertheless, the employment of such a physical solution may have the following implications and prospects: 
\begin{enumerate}
\item
Able to achieve computational reduction (need to generalize to $N$-person problem),
\item
Able to make real-time decisions at the time of information updates, e.g., parameter changes during calculations (flexibility),
\item
Able to interact with information between different species,
\item
Able to interlock the distribution motion directly from physical calculations (water distribution, oil distribution), and 
\item
Eliminate artificiality by using natural phenomena, thus making it easier for people to understand and to reach consensus (compromise).
\end{enumerate}

Real-world problems, such as the distance between two people in a taxi problem, must be addressed often without complete information available. 
We should be flexible and able to respond flexibly if the information is updated along the way. 
This system will become a candidate for overcoming the computational difficulties in such problems.

\section*{Acknowledgements}
This work was supported by the research grant SP001 from SOBIN Institute, LLC.
We would like to thank Dr. Emanuel Pastreich (The Asia Institute, South Korea) and Dr. Alexander Krabbe (The Asia Institute, Germany) for fruitful discussions on this work.

\section*{Author Information}
Correspondence should be addressed to S.-J.K. (kim@sobin.org).

\newpage

\begin{table}
\caption{A bankruptcy Problem in the Talmud.} 
\label{table:1} 
\begin{center}
\begin{tabular}{|c|c|c|c|} \hline 
    & $D_1=100$ & $D_2=200$ & $D_3=300$ \\ \hline
$M=100$ &  33 + 1/3 & 33 + 1/3 &  33 + 1/3 \\ \hline
$M=200$ & 50 & 75 & 75 \\ \hline
$M=300$ & 50 & 100 & 150 \\
\hline 
 \end{tabular}
\end{center}
\end{table}

\begin{figure}[H]
\begin{center}
\label{fig:3}
\includegraphics[width=90mm]{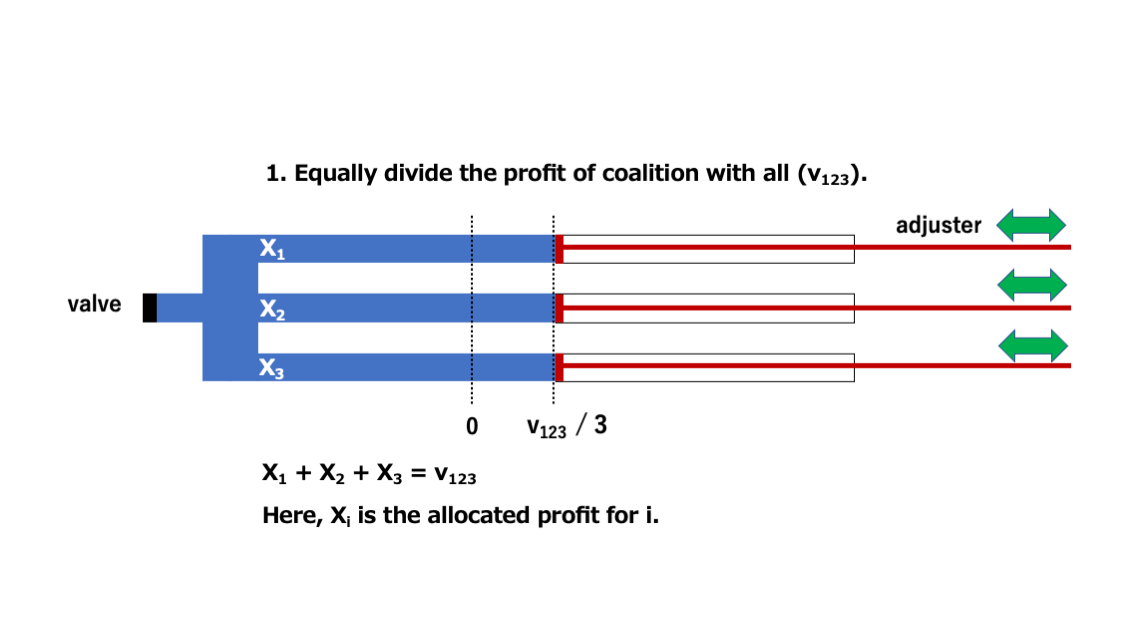}\\
\includegraphics[width=90mm]{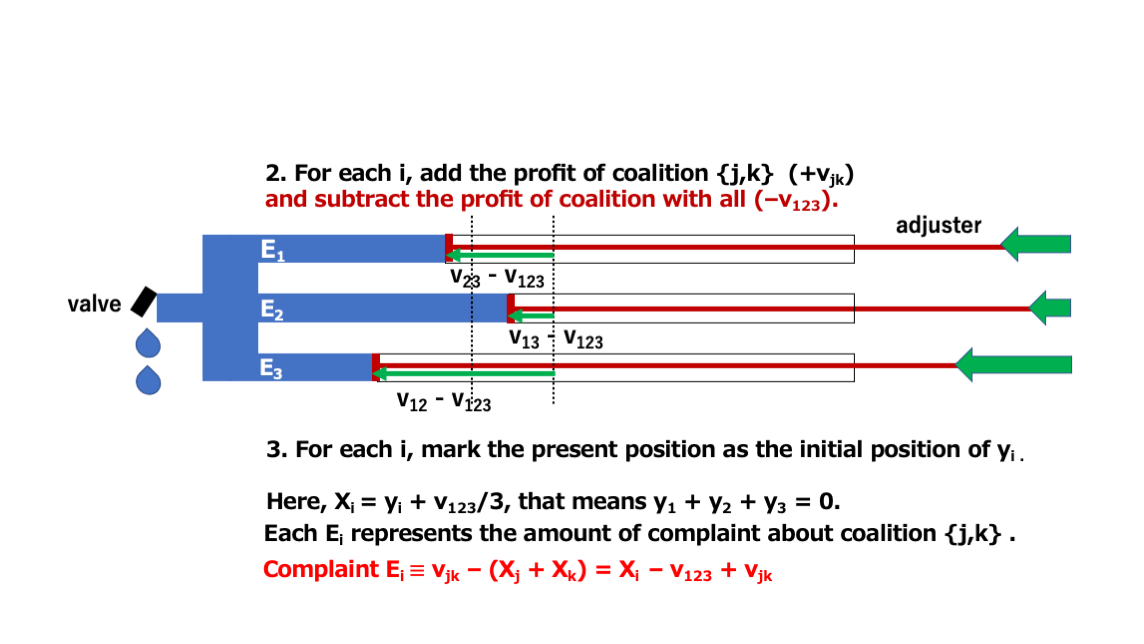}\\
\includegraphics[width=90mm]{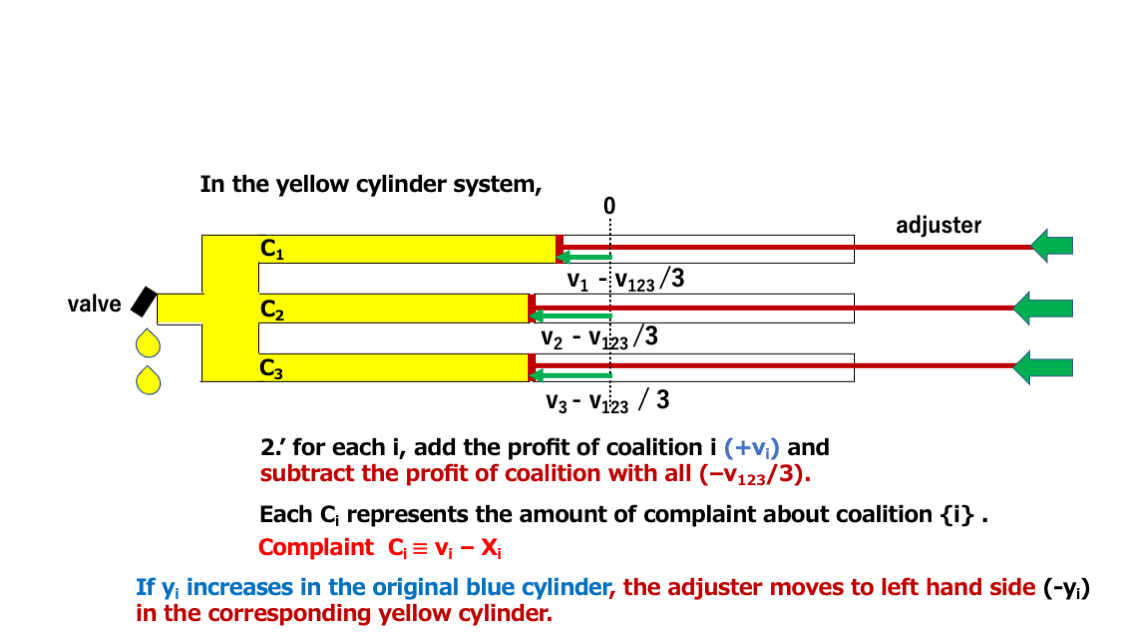}\\
\includegraphics[width=90mm]{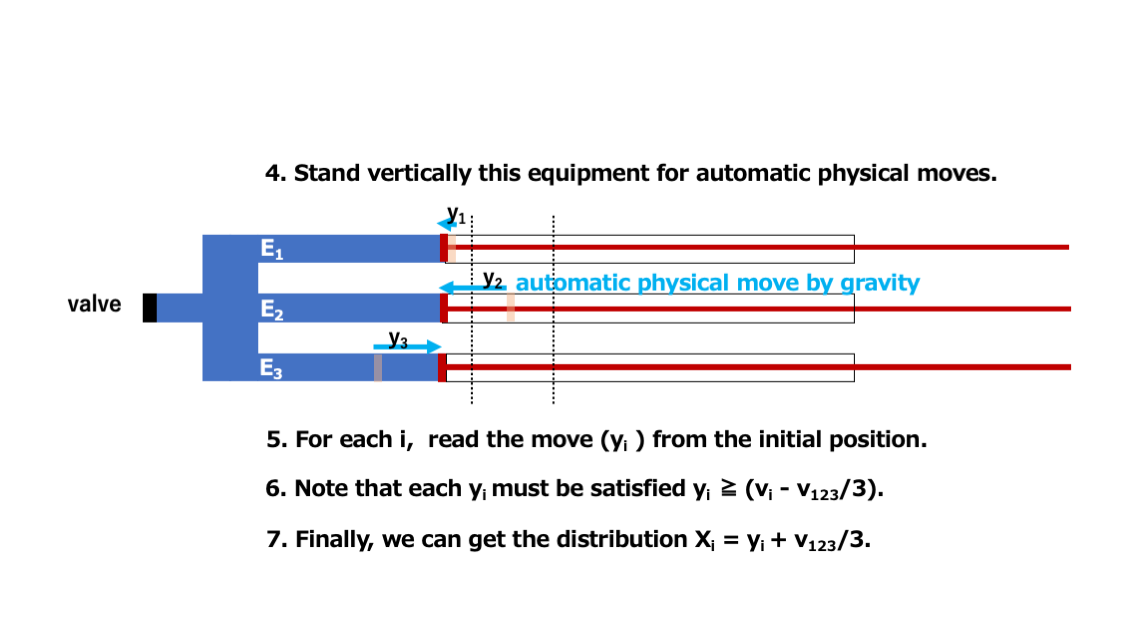}
\caption{Fair distribution method utilising physics in game of characteristic function form.} 
\end{center}
\end{figure}

\begin{figure}[H]
\begin{center}
\label{fig:4}
\includegraphics[width=90mm]{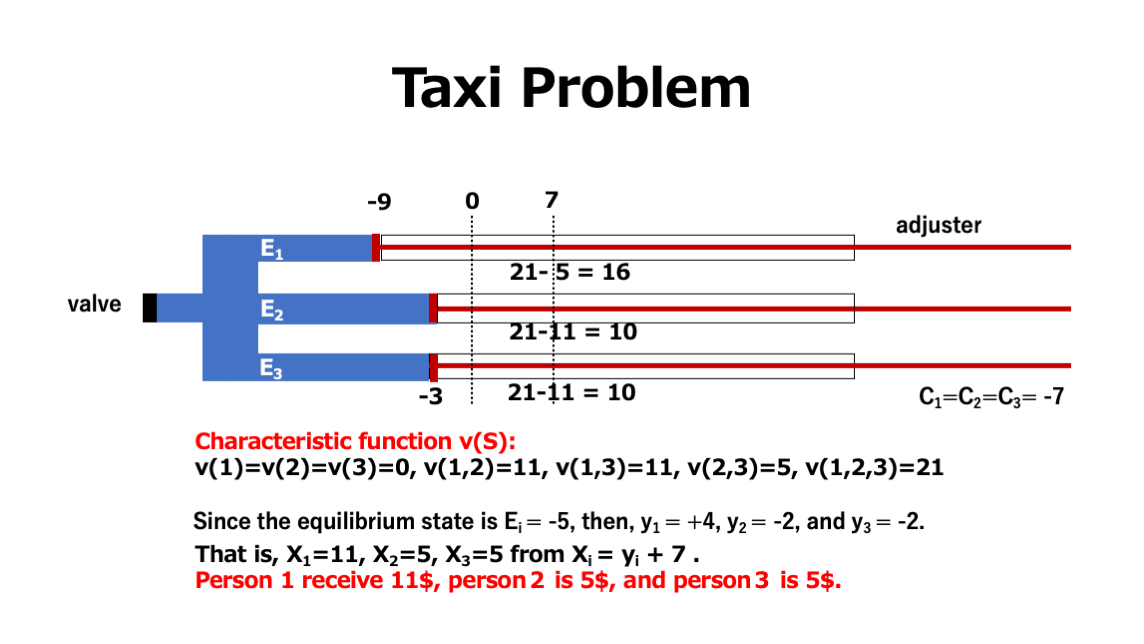}\\
\includegraphics[width=90mm]{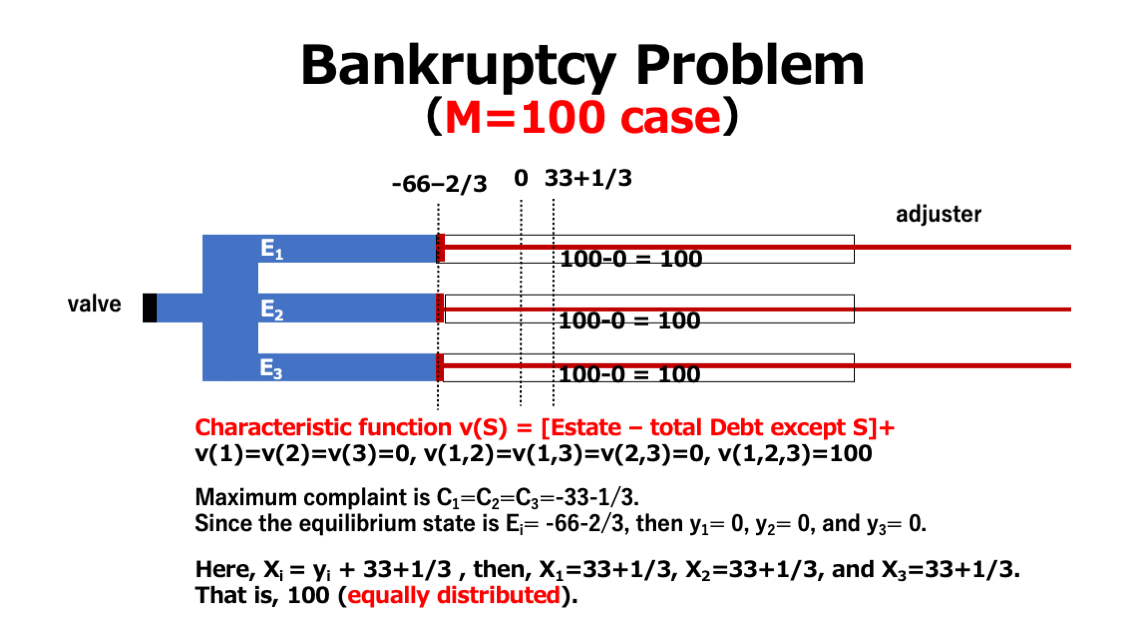}\\
\includegraphics[width=90mm]{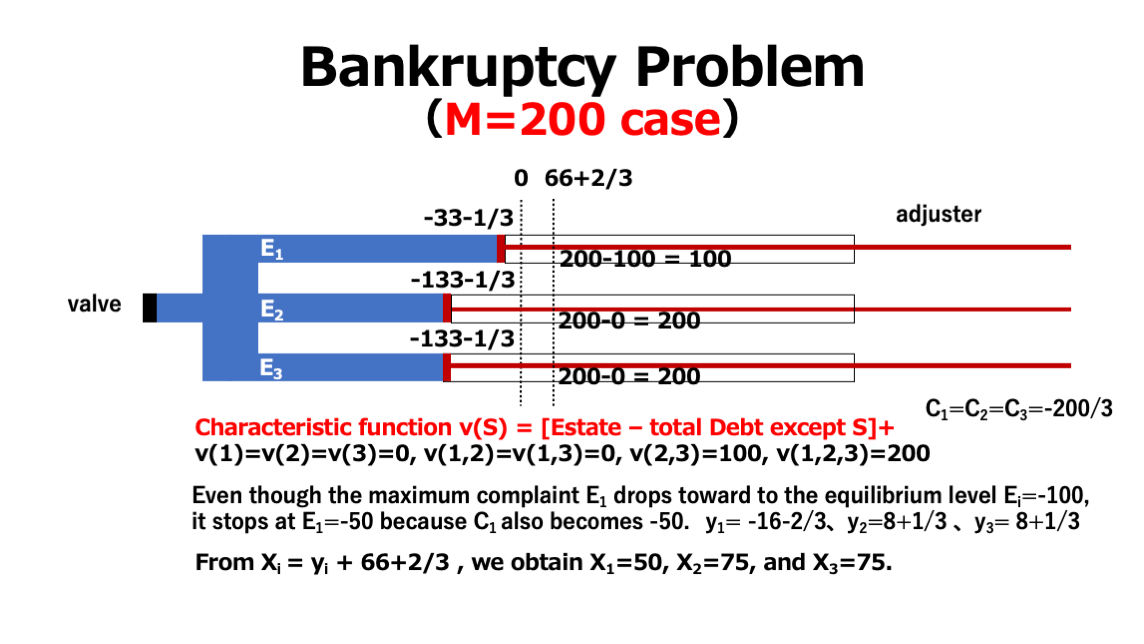}\\
\includegraphics[width=90mm]{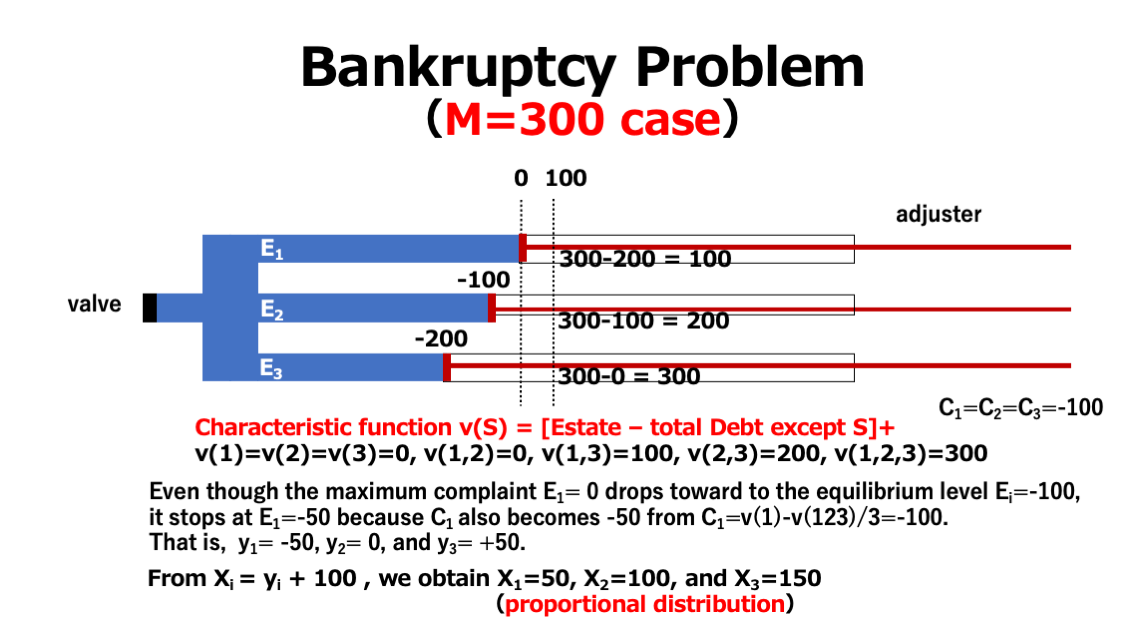} 
\caption{Physical solutions for profit distribution problems in game of characteristic function form.} 
\end{center}
\end{figure}

\newpage

\end{document}